\documentclass[journal,twoside]{IEEEtran}
\usepackage{cite}
\usepackage{amsmath,amssymb,amsfonts}
\usepackage{graphicx}
\usepackage{textcomp}

\usepackage{subfigure}

\usepackage{epsfig}

\usepackage{amssymb}
\usepackage{amsmath}

\usepackage{mathrsfs}

\usepackage{algorithm}
\usepackage{algpseudocode}
\usepackage{enumitem} 
\usepackage{multirow} 
\usepackage{makecell} 
\usepackage{booktabs} 
\usepackage{array} 
\usepackage{float}
\usepackage{tabularx}
\usepackage{hyperref}
\usepackage{cleveref}
\usepackage{color}
\usepackage{placeins}
\usepackage{longtable}
\usepackage{comment}
\usepackage{afterpage}

\newtheorem{theorem}{Theorem}[section]      
\newtheorem{definition}{Definition}[section] 
\newtheorem{property}{Property}[section]    
\newtheorem{corollary}{Corollary}[section]   
\newtheorem{proposition}{Proposition}

\crefname{equation}{eq.}{eqs.}
\Crefname{equation}{Equation}{Equations}
\crefname{algorithm}{alg}{algs}
\Crefname{algorithm}{Alg}{Algs}
\crefname{corollary}{Cor}{Cors}
\Crefname{corollary}{Cor}{Cors}

\def\BibTeX{{\rm B\kern-.05em{\sc i\kern-.025em b}\kern-.08em
    T\kern-.1667em\lower.7ex\hbox{E}\kern-.125emX}}
\begin{document}

\title{Critical States Identification in Power System via Lattice Partition and Its Application in Reliability Assessment}

\author{\thanks{This work was supported by the National Key Research and Development Program of China (No. 2023YFA1011302)}
	\IEEEauthorblockN{Han Hu, Wenjie Wan, Feiyu Chen\IEEEauthorrefmark{1}, Xiaoyu Liu, Bo Yu, Kequan Zhao}
	\thanks{Corresponding Author: Feiyu Chen \quad Email: fchen\_cqnu@163.com}
	\thanks{W. Wan, Han Hu, F. Chen, X. Liu, and Bo Yu are with the National Center for Applied Mathematics in Chongqing,  Chongqing Normal University, China.}
	\thanks{K. Zhao is with the School of Mathematical Sciences, Chongqing Normal University, China.}
}
\maketitle

	\begin{abstract}
	With the increasing complexity of power systems, accurately identifying critical states (the states corresponding to minimal cut sets) and assessing system reliability have become crucial tasks. In this paper, a mathematical lattice structure is employed to represent and partition the state space of power system. Based on this structure, a novel recursive method is proposed to efficiently identify critical states by leveraging lattice partitioning and Optimal Power Flow (OPF) calculations. This method not only enables the extension of failure system states, but also calculates the upper and lower bounds of the Loss of Load Probability (LOLP) in a progressively converging manner. Compared to traditional reliability assessment methods such as State Enumeration (SE) and Monte Carlo Simulation (MCS), this approach offers greater accuracy and efficiency.
	Experiments conducted on the RBTS and RTS79 systems demonstrate that the proposed method accurately identifies all critical states up to a preset order, which are high-risk states. The contribution of these critical states to LOLP highlights their significance in the system. Moreover, the proposed method achieves the analytical value with significantly fewer OPF calculations in RBTS system, reaching acceptable precision of LOLP up to 100 times faster than SE in both the RBTS and RTS systems. 
\end{abstract}

\begin{IEEEkeywords}
	Critical states, Lattice partition, Power system reliability, State extension
	
	
\end{IEEEkeywords}

		\section{Introduction}
		\label{sec1}

		With the development of power systems towards ultra-high voltage, long-distance transmission, and large capacity, the economic and social impacts of power outages have become increasingly significant. Precisely identifying minimal cut sets can provide better insights into the power system’s vulnerabilities, which not only helps assess the reliability of existing systems but also provides important theoretical foundations for enhancing system stability and robustness\cite{Billinton1994} \cite{lee2022reliability}.
		
		The minimum cut set method has been widely applied for fault analysis and system optimization. Traditional methods include Fault Tree Analysis (FTA) techniques such as the Fussel-Vesely algorithm and dual tree method, which determine minimum cut sets through Boolean algebra simplification and dual transformation \cite{billinton1970power}. Network-based analysis methods, such as the Max-Flow Min-Cut Theorem, leverage graph theory and network flow theory to identify minimal cut sets\cite{Ford_Fulkerson_1956}\cite{10.5555/1942094}. Additionally, the Minimum Cut Set-Network Equivalence Method simplifies the solution process for complex distribution systems through network reliability equivalence \cite{billinton1992reliability}. However, the above methods either struggle with dynamic scenarios or primarily rely on topological structures, which limits their ability to account for dynamic changes and load fluctuations, leading to either missed important cut sets or redundant results. 
		
		In addition, when combined with Optimal Power Flow (OPF) analysis, Monte Carlo Simulation (MCS) methods and State Enumeration (SE) can also be employed to identify minimal cut sets\cite{Billinton1994}\cite{billinton1992reliability}. These methods can provide more accurate evaluations by considering both system topology and operational constraints. However, MCS for identifying minimum cut sets is inefficient due to its need for extensive simulations and may overlook rare or low-probability cut sets. SE suffers from exponential computational complexity, making it impractical for large-scale systems. To improve that, the method described in \cite{5342441} uses DC-OPF approach to identify minimal cut sets up to a preset order by enumeration and comparison level by level. Besides, to identify system states with high risk factors, \cite{LIU20081019} \cite{6523187} developed fast sort algorithm (FSA) which obtained high-probability system outage states by arranging the availability of components in ascending order, a bi-level model optimization model for risk assessment was introduced in \cite{ding2016bi_level}, which identifies some outage states which maybe of great importance whereas they may not be the states corresponding to the minimal cut sets. But these methods fail to determine some minimal cut sets completely and accurately in large-scale systems. 
		
		Moreover, all the methods mentioned above for identifying the minimum cut sets are all based on the inclusion-exclusion principle when calculating reliability indices. This is computationally intensive and provides only fuzzy upper and lower bounds although it is also a form of state extension. Some other methods have also performed state extension. Clancy proposed classifying the system state space and perform state extension by the classified sets\cite{CLANCY1983101}, Billiton proposed a state extension method based on a tree structure\cite{BILLINTON1998189}\cite{852155}, and a transmission system reliability evaluation method was proposed in \cite{7741639} based on the impact increment, where the effects of higher-order failure states were incorporated into lower-order failures. These methods can all achieve higher indices within a limited number of assessments but they have limited state extension areas, resulting in minimal practical application effect. 
		
		This paper applied a mathematical structure - lattice, to identify critical states, which correspond to states where only components in minimal cut sets are failed. The proposed method partitions the system state space into lattices recursively so that critical system states can be fully identified up to a preset level by successive evaluation and comparison with less OPF calculations. Meanwhile, the lattice structure can also serve as a state extension tool, ultimately providing more accurate LOLP (Loss of Load Probability) within a limited number of assessments. And the method proposed in this paper overcomes limitation of limited extension range in \cite{BILLINTON1998189} through lattice partition and provides a new and effective solution for power system reliability assessment.
		
		This paper is organized as follows:
		section 2 illustrates the mathematical representation of system state and defines the critical states;
		section 3 gives the definition of lattice in the state space and proposes an approach to partition the state space into lattices by the 1-level states and 2-level states;
		section 4 gives the algorithm for identifying critical states in a 1-normal lattice;
		section 5 proposes the method to determine the critical states and calculating LOLP index of a system;
		section 6 conducts experiments on the RBTS and RTS-79 systems, proving the efficiency of the method.

		\section{Structure Analysis of State Space of Power System}
		\subsection{Basic Concepts of the Power System States}
		Assume a composite power system contains $n$ components numbered by series number $1,2,...,n$, a component is considered either operational or failed in our study. Then a system state is represented as the set of failed components. For example, a system state $s \in S$ with $k$ failed components is represented as 
		\begin{equation}
			s = \{i_1, i_2, \ldots , i_k\}
		\end{equation}
		where components $i_{1}, i_{2}, \ldots, i_{k}$ are failed, and the others are operational. In particular, $\hat{0}=\{ \}$ represents the system state where all components are operational, and $\hat{1}=\{ 1,2,...,n\}$ represents the state in which all components are failed. The term state as used in the following discussion will imply system state. And the state space of the power system is denoted by $S$, which contains $2^n$ states.

		\begin{definition}
			A \(k\)-level state is defined as the state with k failed components.
		\end{definition}
		The level of $s$ is denoted as $\rho(s) $. In particular, $\hat{0}=\{ \}$ is a 0-level state, $\hat{1}=\{ 1,2,...,n\}$ is a $n$-level state. For example, for a 5-component system, $s=\{1,3,5\}$ is a 3-level state where components 1, 3, 5 are failed, and components 2, 4 are operational, and $\rho(s)=3$.

		\begin{definition}
			Failure states are defined as states where load shedding occurs, normal states are defined as states where no load shedding occurs. The terms failure and normal are referred to as the statuses of a state.
		\end{definition}

		\begin{definition}
			Define a function on the system state space as $\Phi :S \to\{0,1\}$
			\begin{equation}
				\Phi(s) = 
				\begin{cases} 
					0 &  s \text{ is a normal state} \\
					1 &  s \text{ is a failure state}
				\end{cases}
			\end{equation}
		\end{definition}
		
		\begin{definition}
			The failure and the normal state space of $S$ are defined as \(F\) and \(N\):				
			\begin{equation}
				F = \{s \in S \mid \Phi(s) = 1\}; \quad N = \{s \in S \mid \Phi(s) = 0\}
			\end{equation}
		\end{definition}
		
		For convenience, we denote the set of all the $k$-level states in \(T \subseteq S\) as \(T^{(k)}\)  in the following discussion. For example, \(F^{(k)}\) is the set of all the $k$-level failure states in $S$.
		
		\subsection{Critical State}
		
		\begin{definition}
			For any \( s, t \in S \), if \( s \subseteq t \), we say that \( s \) is less than or equal to \( t \), denoted as \( s \leq t \). Equivalently, \( t \) is greater than or equal to \( s \), denoted as \( t \geq s \).
		\end{definition}
		
		For example, \(\{1,3,5\} < \{1,3,4,5\}\) and \(\{2,3,5\} > \{3,5\}\). In fact, the relation "\( \leq \)" in \(S\) is a partial order as defined in \cite{sagan2020combinatorics}. Therefore, we can establish the following proposition:
		\begin{proposition}
			\((S, \leq)\) is a partially ordered set, where for any \( s \in S \),
			\begin{equation}
				\{\} = \hat{0} \leq s \leq \hat{1} = \{1, 2, \dots, n\}
			\end{equation}
			Here, \(\hat{0} = \{\}\) is the minimum element of \( S \) and \(\hat{1} = \{1, 2, \dots, n\}\) is the maximum element of \( S \).
		\end{proposition}

		System coherence is the basic assumption of reliability evaluation of composite power system. System coherence implies that the system performance could not get better if an operational component fails, and not get worse if a failed component is repaired. This property is true in most complex power systems, especially in high-level failure state. \cite{BILLINTON1998189}. On this assumption, the following proposition can be established.
		
		\begin{proposition}
			\label{pro:5}
			Assume $s \in S$ and $t \leq s$. Then:
			\begin{enumerate}
				\item If $\Phi(t) = 1$, then $\Phi(s) = 1$.
				\item If $\Phi(s) = 0$, then $\Phi(t) = 0$.
			\end{enumerate}
		\end{proposition}
		The first case in Proposition \ref{pro:5} refers to the extension of failure status from state t to s, we call it state extension, which is consistent with the concept in \cite{852155}.
		
		\begin{definition}
			\label{def:CS}
			A state $c \in S$ is defined as a \textit{critical state} if and only if \(\Phi(c)=1\) and \( \Phi(s) = 0 \) for all \( s < c \).
		\end{definition}
		The set of all the critical states is denoted as $C$. According to Proposition\ref{pro:5}, states greater than a critical state are definitely failure states. In fact, a failure state is either a critical state or is greater than a critical state, then the following proposition is established.
		
		\begin{proposition}
			\label{proposition3}
			\begin{equation}
				C \cup U(C) = F
			\end{equation}
			where 
			\begin{equation}
				U(C)=\{s \in S \mid s>c, \exists c \in C\}  
			\end{equation}
		\end{proposition}
		
		In critical states, the failure of an additional component leads to system outage, whereas repairing a failed component restores the system to normal operation. Specifically, A critical system state refers to the condition where all components in the corresponding minimal cut set have failed. In Section3,  we introduce an important concept of boolean lattice, which helps with critical states identification and reliability calculation.

		\subsection{Boolean Lattice Representation of System State Space}

		\begin{definition}
			A lower bound for \(s,t \in S\) is an element \(l \in S\) such that \(l \leq s\) and \(l \leq t\); and an upper bound for \(s,t \in S\) is an element \(u \in S\) such that \(u \geq s\) and \(u \geq t\).
		\end{definition}

		\begin{definition}
			$s,t$ have a least upper bound or join if there is a state in $S$, denoted $s \vee t$, which is an upper bound for $s,t$ and $s \vee t \leq  v$ for all upper bounds $u$ of $s$ and $t$. Correspondingly, the greatest lower bound or meet of $s$ and $t$ is denoted as $s \wedge t$. 
		\end{definition}
		
		The least upper bounds of all the states in a set $T$ are denoted as $\vee(T)$. For example, \(\{1,3,4,5\} \wedge \{2,3,5\} = \{3,5\}\), \(\{2,3,5\} \vee \{1,3,5\} = \{1,2,3,5\}\),\\ \(\vee(\{  \{1,2\}, \{1,2,5\},\{4,5\} \}) =\{1,2,4,5\} \).

		\begin{definition}
			A boolean lattice $\mathcal{L}$ is a partially ordered set in which every element $x \in \mathcal{L}$ has a unique complement $\bar{x} \in \mathcal{L}$ such that $x \wedge \bar{x} = \hat{0}_{\mathcal{L}}$ and $x \vee \bar{x} = \hat{1}_{\mathcal{L}}$, where $\hat{0}_{\mathcal{L}}$ is the minimal element and $\hat{1}_{\mathcal{L}}$ is the maximal element of $\mathcal{L}$.
		\end{definition}

		\begin{definition}
			Assume $\bar{s}, \underline{s} \in S $, $\underline{s}\leq\bar{s}$, then the corresponding closed interval is defined as \( [\underline{s}, \bar{s}] \):
			\[[\underline{s}, \bar{s}] = \{ s \in S \mid \underline{s} \leq s \leq \bar{s} \}.\] 
			where \( \underline{s} \) is the minimum element and \( \bar{s} \) is the maximum element of the interval $[\underline{s}, \bar{s}]$.
		\end{definition}

		\begin{proposition}
			\label{interval}
			An interval $[\underline{s}, \bar{s}]$ is a boolean lattice. 
		\end{proposition}
		
		Boolean lattice is presented by interval in this paper. In particular, $S$ is an n-dimensional boolean lattice presented as \([\hat{0}, \hat{1}]\).
		
		\begin{definition}
			Given a boolean lattice $\mathcal{L}=[\underline{s},\bar{s}] \subseteq S$, the dimension of $\mathcal{L}$ is defined as 
			\[D(\mathcal{L})=\rho (\bar{s} )-\rho (\underline{s})\]
			and $\mathcal{L}$ can be called a $D(\mathcal{L})$-dimension boolean lattice.

			In particular, if $\underline{s}=\bar{s}$, then $[\underline{s},\bar{s}] = \{\underline{s}\}$, which is a 0-dimensional boolean lattice and contains only one state. 
		\end{definition}    
	
		The term lattice as used in further discussions will imply boolean lattice. Two special types of lattice are introduced as follows. 
		
		\begin{definition}
			A $k$-normal lattice is defined as a lattice in which all $k$-level states are normal.
		\end{definition}

		\begin{definition}
			\label{def:1}
			A failure lattice is defined as a lattice whose minimal element is a failure state.
		\end{definition}
		
		It is noted that a $1$-normal lattice is also a $0$-normal lattice according to Proposition\ref{pro:5}. All states in a failure lattice are failure states, since they are greater than or equal to the minimal failure state, therefore, failure lattice cannot be a $k$-normal lattice. Based on the fact that any state strictly less than a critical state must be a normal state, an important property is given as follows. 
		\begin{property}
			\label{property}
			In a failure lattice, only the minimal element may be a critical state.
		\end{property}
		
		According to Property\ref{property}, if failure lattices can be partitioned out from S, the OPF calculations and comparisons between states will be reduced when searching for critical states, since only the minimal elements of the failure lattices need to be considered. Thus, Section 3 proposes a basic technique to partition out failure lattices from the $S$.

		\section{Lattice Partition of State Space of Power system}
		Before illustrating the lattice partition, we introduce some notations in a \(n\)-dimension lattice \(\mathcal{L}\). 
		
		Let \(\mathcal{L}\) be the state space of an \(n\)-component system state space with ordered 1-level states, the 1-level states and 2-level states of \(\mathcal{L}\) are denoted as 
		
		\begin{equation}
			\begin{aligned}
				a_i \quad = & \quad \{i\} \qquad \text{for }  1 \le i \le n \\
				a_{ij} \quad = & \quad \{i,j\} \quad \text{for } 1 \leq i < j \leq n
			\end{aligned}
		\end{equation}

		The corresponding conjugate state of $a_i$ and $a_{ij}$ are denotes as:
		\begin{equation}
			\begin{aligned}
			t_i \quad = & \quad \vee(\{a_k\}_{k=i}^n) \qquad \text{for }  1 \le i \le n \\
			t_{ij} \quad = & \quad \vee(\{a_{ik}\}_{k=j}^{n}\}) \quad \text{for } 1 \leq i < j \leq n
			\end{aligned}
		\end{equation}
		
		In particular, we set \(a_0 = t_{n+1}=\hat{0}_{\mathcal{L}}, t_{i(n+1)}=a_i\). 
		
		For a 5-dimension lattice, the above notations are shown in Fig.\ref{fig1a}.

		\subsection{Partition of Lattice by 1-level states}                		
		\begin{theorem}
			\label{thm_partition1}
			Let \(\mathcal{L}\) be a n-dimensional lattice with ordered 1-level states \( \{a_{1},a_{2},...,a_{n}\} \), then $\mathcal{L}$ can be partitioned into $m+1$ lattices:
			\begin{equation}                
				\label{eq_partition1}
				\mathcal{L} =  \bigcup_{i=1}^{m} \overline{\mathcal{L}}_i \cup \underline{\mathcal{L}}_m \quad \text{for } 1 \le m \le n
			\end{equation}    
			where 
			\begin{equation}
				\overline{\mathcal{L}}_i=[a_i, t_i], \quad  \underline{\mathcal{L}}_m=[\hat{0}_\mathcal{L},t_{m+1}]
			\end{equation}
		\end{theorem}
		
		The partition result is not unique due to the different order of 1-level states. 
		
		\begin{corollary}
			\label{coro1_partition1}
			If the 1-level states are ordered such that \( \{a_{1},a_{2},...,a_{f}\} \) are all the 1-level failure states, then each \( \overline{\mathcal{L}}_i\) in (\ref{eq_partition1}) is a failure lattice, and  \( \underline{\mathcal{L}}_m\) in (\ref{eq_partition1}) is a 1-normal lattice.
		\end{corollary}
		
		If \(\mathcal{L}\)  is a n-dimensional lattice containing $f<n$ failure 1-level states, then $\mathcal{L}$ can be partitioned into $f$ failure lattices and one 1-normal lattice according to Corollary\ref{coro1_partition1}.

		

		\subsection{Partition of Lattice by 1-level and 2-level states}
		\begin{theorem}
			\label{thm_partition2}
			Let \(\mathcal{L}\) be a n-dimensional 1-normal lattice ($n \ge 2$) with ordered 1-level states \(\{a_1, a_2,...,a_n\}\), $\mathcal{L}$ can be partitioned as follows:

			\begin{equation}
				\label{eq_partition2}
				\begin{aligned}
					\mathcal{L} \quad =& \quad \bigcup_{i=1}^{n-1} \overline{\mathcal{L}}_i \cup \underline{\mathcal{L}}_{n-1} \\
					\quad =& \quad \bigcup_{i=1}^{n-1} [ \bigcup_{j=i+1}^{f_i} \overline{\mathcal{L}}_{ij}
					\cup  \underline{\mathcal{L}}_{if_i} ] \cup \underline{\mathcal{L}}_{n-1} \\
					\quad =& \quad \bigcup_{i=1}^{n-1} \bigcup_{j=i+1}^{f_i} \overline{\mathcal{L}}_{ij} \cup \bigcup_{i=1}^{n-1}   \underline{\mathcal{L}}_{if_i} \cup \underline{\mathcal{L}}_{n-1}\\
					& \text{for } i+1 \leq f_i \leq n
				\end{aligned}
			\end{equation}

			where 
			\begin{eqnarray}
				&\overline{\mathcal{L}}_i=[a_i, t_i], \quad \quad \underline{\mathcal{L}}_{i}=[\hat{0}_\mathcal{L},t_{i+1}] \\
				&\overline{\mathcal{L}}_{ij}=[a_{ij}, t_{ij}], \quad  \underline{\mathcal{L}}_{ij}= [a_i, t_{i(j+1)}]
			\end{eqnarray}                	
		\end{theorem}

		
		\begin{figure*}[htbp]
			\subfigure[]{
				\begin{minipage}[t]{0.3\linewidth}
					\centering
					\includegraphics[width=\textwidth]{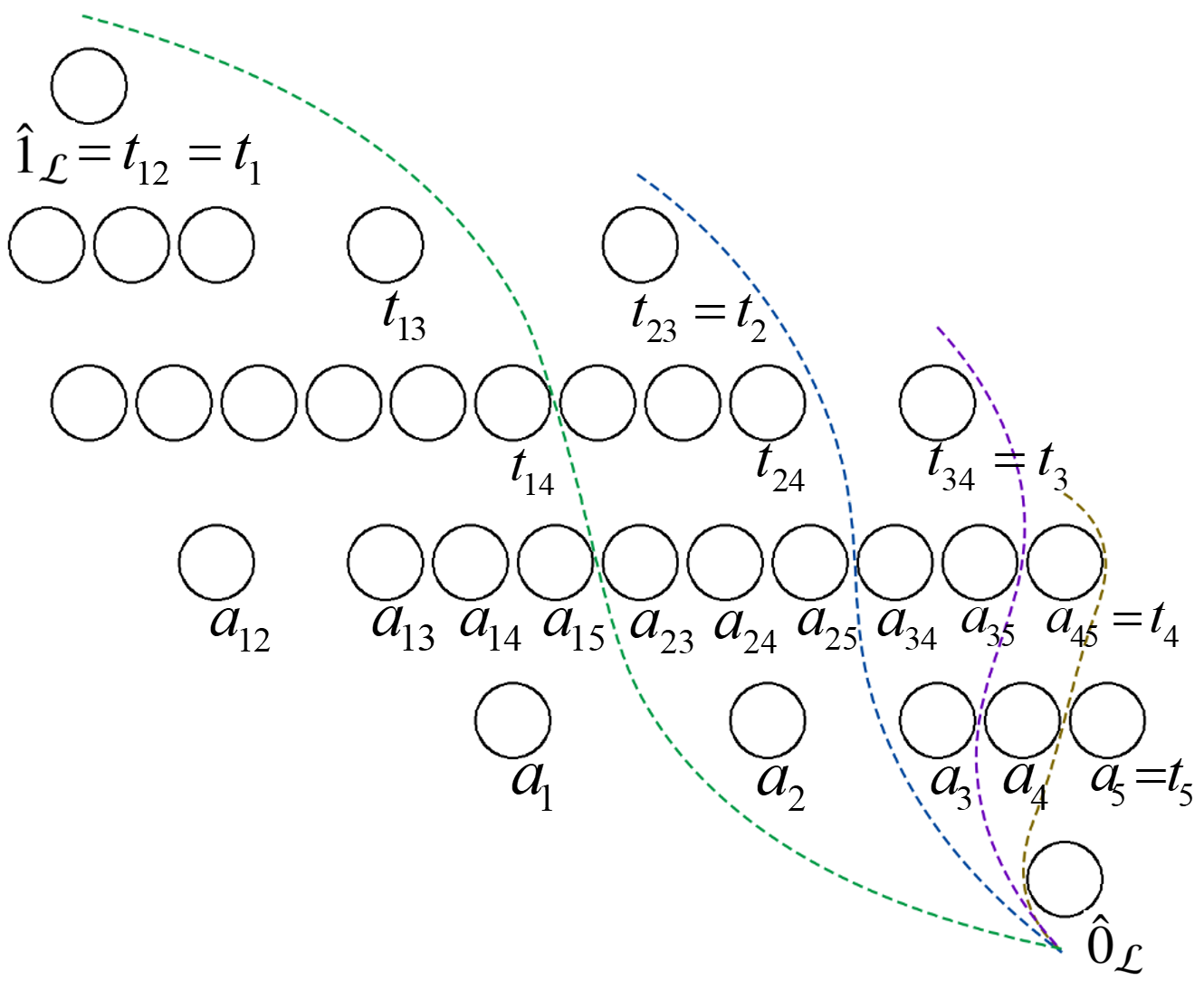}
					\label{fig1a}
				\end{minipage}
			}%
			\hspace{0.03\linewidth} 
			\subfigure[]{
				\begin{minipage}[t]{0.3\linewidth}
					\centering
					\includegraphics[width=\textwidth]{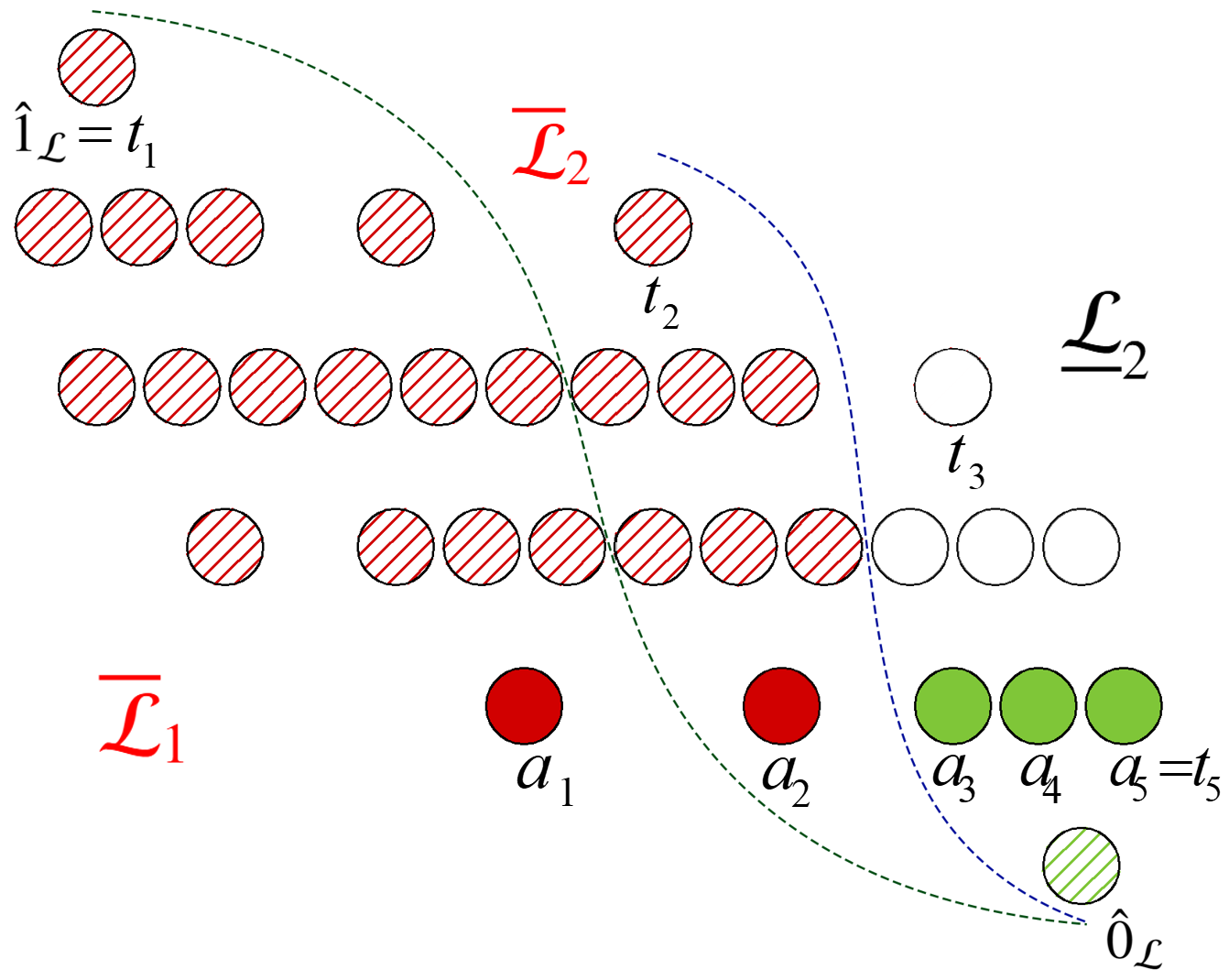}
					\label{fig1b}
				\end{minipage}
			}
			\hspace{0.03\linewidth} 
			\subfigure[]{
				\begin{minipage}[t]{0.3\linewidth}
					\centering
					\includegraphics[width=\textwidth]{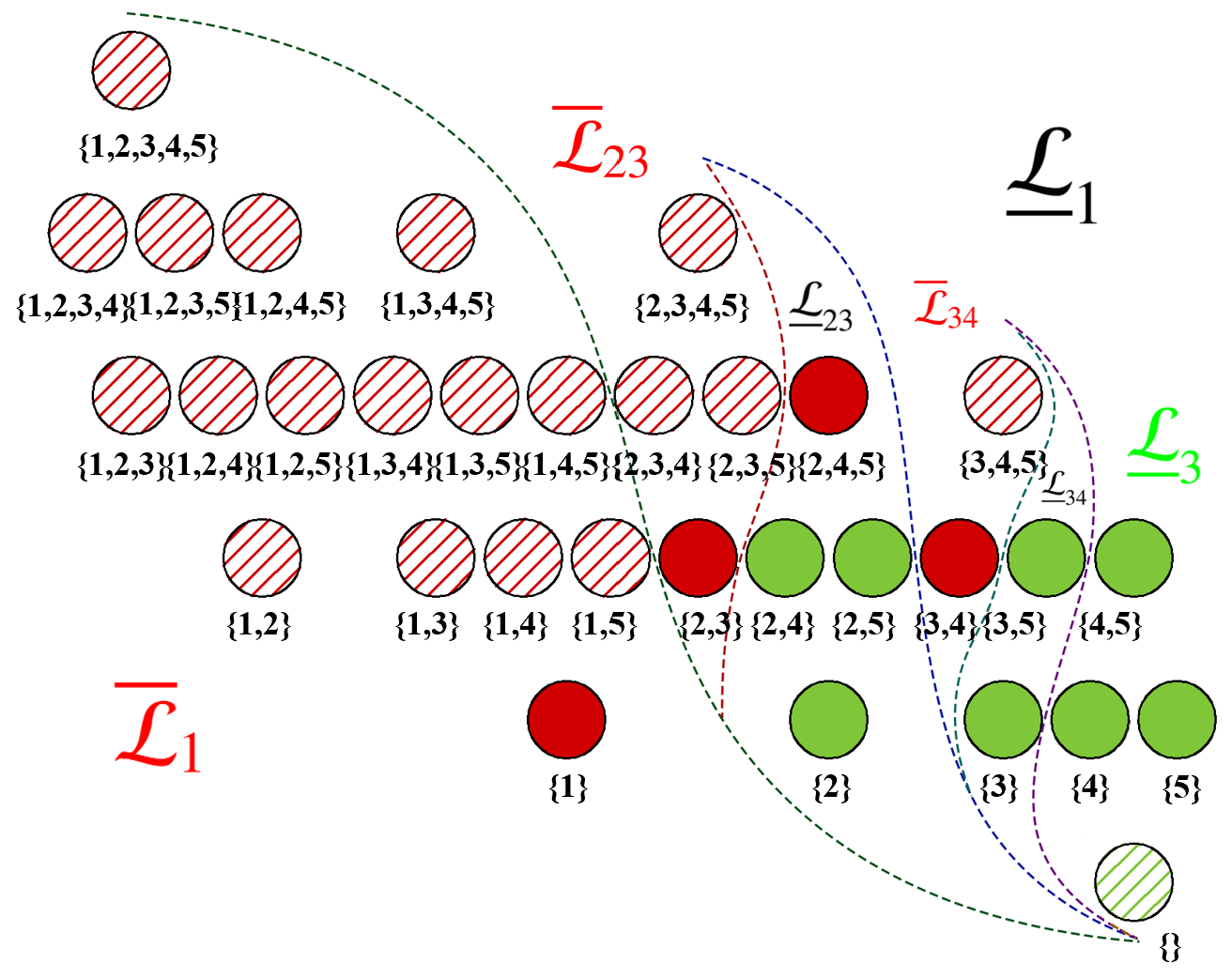}
					\label{fig1c}
				\end{minipage}
			}
			\caption{Labeling and examples of partition of a 5-dimension lattice}
			\label{fig1}
		\end{figure*}

		\begin{corollary}
			\label{coro1_partition2}
			If the 1-level states are ordered such that \(\{a_{i i+1},a_{i i+2},\dots, a_{i f_i}\}\) for \(i=1,2,\dots,n-1\) are all the failure 2-level states, then each \(\overline{\mathcal{L}}_{ij}\) in (\ref{eq_partition2}) is a failure lattice.
		\end{corollary}
		
		
		If \(\mathcal{L}\) is a n-dimensional 1-normal lattice containing $f$ 2-level failure states, then $\mathcal{L}$ can be partitioned into $f$ failure lattices, $n-2$ 1-normal lattice, and one normal lattice according to Corollary\ref{coro1_partition2}.

		For example, consider the 5-dimension lattice \(\mathcal{L}=[\{\},\{1,2,3,4,5\}]\) with ordered 1-level states \(\{\{1\},\{2\},\{3\},\{4\},\{5\}\}\), assume that 1-level states \(\{1\}\) and \(\{2\}\) are failure, then \(\mathcal{L}\) can be partitioned into two failure lattices \(\overline{\mathcal{L}}_1\) and \(\overline{\mathcal{L}}_2\) and one 1-normal lattice \(\underline{\mathcal{L}}_{2}\) based on Corollary\ref{coro1_partition1}:
		
	\begin{equation}
		\begin{aligned}
			\mathcal{L} \quad =& \quad    \textcolor{red}{\overline{\mathcal{L}}_1} \cup \textcolor{red}{\overline{\mathcal{L}}_2} \cup \underline{\mathcal{L}}_{2}  \\
			\quad =& \quad   \textcolor{red}{[\{1\},\{1,2,3,4,5\}]} \cup \textcolor{red}{[\{2\},\{2,3,4,5\}]}  \\
			 \quad & \quad  \cup [\{\},\{3,4,5\}]
		\end{aligned}
	\end{equation}

		which is shown in Fig.\ref{fig1b}, where red circles represents failure state and green circles represents normal states. In another case, if there is only one 1-level failure state \(\{1\}\) in \(\mathcal{L}\), then \(\mathcal{L}\) can be partitioned into \(\overline{\mathcal{L}_1}\) and \(\underline{\mathcal{L}}_{1}\), and $\underline{\mathcal{L}}_{1}=[\{\},\{2,3,4,5\}]$ is a 4-dimension 1-normal lattice with ordered 1-level states \(a_i=\{i+1\}\) for \(i=1,2,\dots,4\), we assume it contains two 2-level failure states \(\{2,3\}\) and \(\{3,4\}\). Then $\underline{\mathcal{L}}_{1}$ can be partitioned into two failure lattices, two 1-normal lattices and one normal lattice based on Corollary\ref{coro1_partition2}:

		\begin{equation}
		\begin{aligned} 
			\mathcal{L} \quad =& \quad   \textcolor{red}{\overline{\mathcal{L}}_1} \cup \underline{\mathcal{L}}_{1} \nonumber \\
			\quad =& \quad   \textcolor{red}{\overline{\mathcal{L}}_1}  \cup \textcolor{red}{\bigcup_{i=2}^{3} \bigcup_{j=i+1}^{f_i} \overline{\mathcal{L}}_{ij}} \cup \bigcup_{i=2}^{3}   \underline{\mathcal{L}}_{if_i} \cup \textcolor{green}{\underline{\mathcal{L}}_{3}} \nonumber \\
			\quad =& \quad  \textcolor{red}{\overline{\mathcal{L}}_1} \cup \textcolor{red}{\overline{\mathcal{L}}_{23}} \cup \textcolor{red}{\overline{\mathcal{L}}_{34}} \cup \underline{\mathcal{L}}_{23} \cup \underline{\mathcal{L}}_{34} \cup \textcolor{green}{\underline{\mathcal{L}}_{3}} \\
			\quad =& \quad   \textcolor{red}{[\{1\},\{1,2,3,4,5\}]} \cup \textcolor{red}{[\{2,3\},\{2,3,4,5\}]} \\
		    & \quad \cup \textcolor{red}{[\{3,4\},\{3,4,5\}]} \cup  [\{2\},\{2,4,5\}] \\
			& \quad \cup [\{3\},\{3,5\}] \cup \textcolor{green}{[\{\},\{4,5\}]}
		\end{aligned}
		\end{equation} 
		   
		which is shown in Fig.\ref{fig1c}. 
		

		
		

		\section{Analysis of 1-normal Lattice}
		
		According to the above theorems and corollaries, a 1-normal lattice can be divided into several failure sublattices and one lower-dimensional 1-normal sublattice depending on the status of 2-level states. 
		
		In general, the status of states are identified by OPF calculation, which is time-consuming for large-scale systems. Fortunately, the use of critical states, which are the minimal elements of failure state space, is benefiting to reducing the number of OPF calculations.                 
		\begin{equation}\label{judgment of failure state}
			s \in F  \quad \quad \text{~if~}   \Phi(s) = 1 \text{~~~or} \quad  s \in U(\hat{C})  
		\end{equation}      
		where $\hat{C}$ is the current set of critical states.
		
		The critical states should be identified by OPF calculations and level-by-level comparisons throughout the system space by the definition. In fact, comparing with the current set of critical states is sufficient for critical state identification if the set of critical states is updated level-by-level.
		\begin{equation}\label{judgment of critical state}
			s \in C  \quad \quad \text{~if~}   \Phi(s) = 1  \text{~~and} \quad  s \notin U(\hat{C})
		\end{equation}
		
		Based on formulas (\ref{judgment of failure state}) and (\ref{judgment of critical state}), we propose an efficient strategy for the identification of 2-level failure states and 2-level critical states in a 1-normal lattice.  
		
		Step 1: Judge the relationship between a 2-level state \(s\) and the current set of critical states \(\hat{C}\),
		\begin{equation}
			s \in F  \quad \text{~but~}  \quad s \notin C; \quad \quad \text{~if~} s \in U(\hat{C})
		\end{equation}                      
		
		Step 2: Evaluate the status of $s$ by OPF calculation if $s \notin U(\hat{C})$,
		\begin{equation}
			\begin{aligned}
				s \in F  \quad \text{~and~}  \quad s \in C; \quad \quad \text{~if~} \quad \Phi(s) = 1  \\
				s \in N  \quad \text{~and~}  \quad s \notin C; \quad \quad \text{~if~} \quad \Phi(s) = 0 
			\end{aligned}
		\end{equation}
		
		After repeating the above two steps until the 2-level space $\mathcal{L}^{(2)}$ is traversed, all the 2-level failed states of the 1-normal lattice $\mathcal{L}$ are identified.

		Denote $f_i$ as the number of 2-level failed states of $\mathcal{L}$ containing in $\overline{\mathcal{L}}_i$, and $f = \sum_{i=1}^n f_i$ is the total number of 2-level failed states in $\mathcal{L}$. Clearly, $f_i$ varies for different orders of components. 
		
		In order to partition out the failure lattices with a larger probability as much as possible, we select the order such that
		\begin{equation}\label{condition of order}
			f_i \ge f_j  \quad \quad \text{~for all~} \quad i > j
		\end{equation} 
		
		The 1-normal lattice $\mathcal{L}$ can be partitioned according to Corollary\ref{coro1_partition2}, and the set of critical states is updated simultaneously. The whole procedure is described in the following algorithm:               
		
		\begin{algorithm}[H]
			\caption{1-normal Lattice Analysis Algorithm}
			\label{1-normal lattice analysis}
			\begin{algorithmic}[1]
				\Statex \textbf{Input:} A 1-normal lattice $\mathcal{L}$, the current set of critical states \(\hat{C} \), current OPF number $N$ (if needed)
				\Statex \textbf{Output:} Updated \(\hat{C}\), partition of \(\mathcal{L}\) (may include failure lattices and a lower-dimension 1-normal lattice in \(\mathcal{L}\))
				\Statex \textbf{Begin}
				\State Set $\Phi(s) = 0$ for all $s \in \mathcal{L}^{(2)}$
				\For{$s \in \mathcal{L}^{(2)}$}
				\If{\(s \in U(\hat{C})\)}
				\State $\Phi(s) = 1$
				\Else
				\State Evaluate $s$ by OPF
				\State \(N = N+1\)
				\If{$\Phi(s) = 1$}
				\State $\hat{C} = \hat{C} \cup \{s\}$
				\EndIf
				\EndIf
				\EndFor
				\State Select the order such that (\ref{condition of order}) holds
				\State Partition \(\mathcal{L}\) by (\ref{eq_partition2}) with the above order
				\Statex \textbf{End}		
			\end{algorithmic}
		\end{algorithm}

		\section{Critical States Identification via Lattice Partition}

		Let $S$ be the state space of a given system, the overall procedure will be discussed in this section in order to identify the critical states of $S$ and assessment reliability of the system. The first step is evaluate all the 1-level states in $S$ by OPF and update \(S\) as follows:
		\begin{equation}
			\label{Update of S}
			S = 
			\begin{cases} 
				S &  \textbf{if} \quad  F^{(1)}=\emptyset \\
				\underline{S}_f &  \textbf{if} \quad F^{(1)}=\{a_1,a_2,...,a_f\}
			\end{cases}
		\end{equation}
		where \(f\) is the number of 1-level failure states of \(S\), and \(\underline{S}_f \) is the 1-normal lattice partitioned from \(S\) using (\ref{eq_partition1}).

		The 1-normal lattice \(\underline{S_f}\) can be partitioned by (\ref{eq_partition2}). Each 1-normal lattice partitioned from $\underline{S}_f$ can also be partitioned further using the method described in Algorithm 1. Then a series of 1-normal lattices in higher-level state space are generated. Thus, the method we will discuss, which we refer to as CSILP(Critical States Identification via Lattice Partition), actually proceeds level by level as the iterative process unfolds.

		\begin{definition}
			Define the probability function on the system state space as $ P:S \to \left [ 0,1 \right ]$
			\begin{equation}
				P(s)=\prod_{i\in s}^{}p_{i}\prod_{i\notin s}^{}q_{i}
			\end{equation}
			where $p_{i}$ is the failure probability of component $i$, and $q_{i}$ is the operational probability of component i, and $p_{i}+q_{i}=1$.
		\end{definition}
		
		\begin{definition}
			Let a lattice be \(\mathcal{L} = [ \underline{s}, \overline{s} ]\), the probability of \(\mathcal{L}\) is defined as sum of probability of all the states in  \(\mathcal{L}\):
			\begin{equation}
				\label{prob of lattice}
				P([ \underline{s}, \overline{s} ]) = \prod_{i \in \underline{s}} p_i \prod_{i \notin \overline{s}} q_i,
			\end{equation}
		\end{definition} 
		
		Since many failure lattices will be partitioned from the original lattice, LOLP of a system can be calculated as the sum of probabilities of all the failure lattices. The lower and upper bounds of LOLP can be calculated as follows:
		\begin{equation}\label{eq:LOLP calculation}
			\underline{\text{LOLP}} = \sum_{L \in F_{\mathcal{L}}} P(\mathcal{L}), \quad \overline{\text{LOLP}} = 1 - \sum_{s \in W^{'}} p(s)
		\end{equation}
		where \( F_{\mathcal{L}} \) is the set of all the failure lattices, \( W^{'} \) is the set of normal states that have been evaluated by OPF, and the value of \underline{\text{LOLP}} is considered the final value of LOLP for the power system assessment.
		
		We set three convergence criteria for the algorithm:
		\begin{enumerate}
			\item OPF number($N$): The number of power flow calculations.
			\item Gap between upper and lower bounds of LOLP ($\delta$): 
			\begin{equation}
				\delta= \overline{\text{LOLP}}-\underline{\text{LOLP}}
				\label{eq:gap of LOLP}
			\end{equation}
			\item Level($k$): Level of states.
		\end{enumerate}
		
		The algorithm is considered to have reached a certain level of precision when $N$ reaches a certain preset threshold $N_*$, or $\delta$ is less than preset gap $\delta_*$, or all the states of the first \(k_*\) level have been searched, with \(k_*\) be the preset maximal level. The three convergence criteria are also applicable to SE method, whereas only the first criterion is applicable to MCS method.

		\makeatletter
		\renewcommand{\Statex}[1][0]{%
			\setlength\@tempdima{\algorithmicindent}%
			\ifnum#1=0\relax
			\item[]%
			\else
			\item[\hskip\dimexpr#1\@tempdima\relax]%
			\fi}
		\makeatother
		
		The whole algorithm for identifying critical states and calculating failure probability of a power system is as follows. 
		
		\begin{algorithm}[htbp]
			\caption{Critical State Identification via Lattice Partition}
			\begin{algorithmic}[1]
				\Statex \textbf{Input:} State space of a \( n \)-component system \(S\), $k_*$(maximum state search level), $N_*$(maximal number of OPF applied to states), ${\delta}_*$(minimal gap between the upper and lower bounds of LOLP)
				\Statex \textbf{Output:}  Critical set \( C \), LOLP
				\Statex \textbf{Begin}
				\State Initialize $k=1$, $\delta=1$, $N=0$, $\hat{C}=\emptyset$
				\State Evaluate all the states in $S^{(1)}$ by OPF  
				\State $N= N+|S^{(1)}|$, $\hat{C} = F^{(1)}$
				\State Update $S$ by (\ref{Update of S})
				\State Initialize 1-normal lattices set \( X = \{S\} \)
				\While{\(N<N_*\) or \(\delta<{\delta}_*\) or  $k< k_*$ }
				\State Apply Algorithm 1 to every lattice in \(X\)
				\State Update $X$ as the union of all the 1-normal lattices with dimension\(\geq 2\) gained in the last step
				\State Calculate \(\underline{\text{LOLP}}\) and \(\overline{\text{LOLP}}\) using  (\ref{eq:LOLP calculation})
				\State  Calculate $\delta$ using (\ref{eq:gap of LOLP})
				\State  $k=k+1$
				\EndWhile
				\State \(C=\hat{C}\), \(\text{LOLP}=\underline{\text{LOLP}}\)
				\State \textbf{End}
			\end{algorithmic}
		\end{algorithm}

		As an example, the above algorithm is run on the synthetic 5-component system shown in Figure2(c). Twelve OPF calculations are performed on the states represented by solid red and green circles, and the following information can be obtained.       
		
		The critical states are represented by solid red circles, and the set of critical states is:
		\begin{equation}
			C=\{\{1\},\{2,3\},\{3,4\},\{2,4,5\}\}
		\end{equation}	
		
		The state space $\mathcal{L}$ can be partitioned into four failure lattices and four normal lattices:
		\begin{equation} 
			\begin{aligned}
			\mathcal{L} \quad =& \quad \textcolor{red}{[\{1\},\{1,2,3,4,5\}]} \cup \textcolor{red}{[\{2,3\},\{2,3,4,5\}]} \\
			\quad \cup& \quad  \textcolor{red}{[\{3,4\},\{3,4,5\}]} \cup  \textcolor{red}{[\{2,4,5\},\{2,4,5\}]}  \\
			\quad \cup& \quad  \textcolor{green}{[\{2,4\},\{2,4\}]} \cup \textcolor{green}{[\{2\},\{2,5\}]}  \\
			\quad \cup& \quad  \textcolor{green}{[\{3\},\{3,5\}]} \cup \textcolor{green}{[\{\},\{4,5\}]}
	    	\end{aligned}
		\end{equation}  
		
		The LOLP index can be calculated analytically:

		\begin{equation} 
			\begin{aligned}
				\text{LOLP} \quad =& \quad P([\{1\},\{1,2,3,4,5\}]) \nonumber \\
				\quad +& \quad P([\{2,3\},\{2,3,4,5\}]) \\
				\quad +& \quad P([\{3,4\},\{3,4,5\}])  \\
				\quad +& \quad  P([\{2,4,5\},\{2,4,5\}]) 
			\end{aligned}
		\end{equation} 
		
		where the formula for calculating the probability of lattice is given in (\ref{prob of lattice}).			
		
		In the above example, the minimum states of failure lattices are all critical states. However, this is not a common scenario. More often, some failure lattices have minimal elements that are not critical states themselves but are greater than a certain critical state. Regardless of the scenario, the failure probability contributed by these failure lattices to the reliability index is essentially derived from certain critical states. 
		In fact, the contribution of a critical state to the reliability index is actually the sum of the probabilities of all failure lattices it covers, which is significantly greater than the failure probability it contributes on its own.

		\section{Numerical Experiments}	
		
		The proposed CSILP method for critical states identification and system reliability assessment was tested on two power systems: a 6-bus system RBTS and a 24-bus system RTS79. The identified critical states for both systems will be presented in order of their levels, along with the proportion of each level's critical states relative to the total number of states at that level. The reliability assessment results, specifically the Loss of Load Probability (LOLP), were compared among Sequential Enumeration (SE) method, Monte Carlo Simulation (MCS) method and the CSILP method. And we set different values for each convergence criterion to compare efficiency and accuracy of the three methods. The experiments were conducted on a standard PC equipped with an Intel(R) Core(TM) i9-14900K CPU at 3.20 GHz and 32GB RAM, using MATLAB R2020b with the Yalmip and Gurobi toolboxes.

		\begin{table}
			\centering
			\caption{Critical States (RBTS)}
			\label{FCS_RBTS} 
			\resizebox{0.9\linewidth}{!}{ 
				\begin{tabular}{ccc} 
					\toprule[1pt] 
					Level & \makecell{Number \\/ Proportion} & Critical State \\ \hline
					1 & 1 / 5\% & \{20\} \\ \hline
					\multirow{2}{*}{2} & \multirow{2}{*}{19 / 10\%} & \{1,2\},\{1,4\},\{1,7\},\{1,8\},\{1,9\},\{1,10\},\{1,11\},\{2,4\},\{2,7\},\{2,8\} \\ 
					~ & ~ & \{2,9\},\{2,10\},\{2,11\},\{4,7\},\{7,8\},\{7,9\},\{7,10\},\{7,11\},\{16,19\} \\ \hline
					\multirow{3}{*}{2} & \multirow{3}{*}{15 / 1.32\%} & \{4,8,9\},\{4,8,10\},\{4,8,11\},\{4,9,10\},\{4,9,11\},\{4,10,11\}, \\ 
					~ & ~ & \{8,9,10\},\{8,9,11\},\{8,10,11\},\{9,10,11\},\{12,13,17\}, \\ 
					~ & ~ & \{12,14,17\},\{12,17,18\},\{13,14,18\},\{14,15,19\} \\ \hline
					\multirow{3}{*}{4} & \multirow{3}{*}{10 / 0.21\%} & \{1,3,5,6\},\{1,14,15,16\},\{2,3,5,6\},\{2,14,15,16\},\{3,5,6,7\}, \\ 
					~ & ~ & \{7,14,15,16\},\{12,15,16,17\},\{12,15,17,19\},\{13,15,16,18\} \\ 
					~ & ~ & \{13,15,18,19\} \\ \hline
					\multirow{5}{*}{5} & \multirow{5}{*}{17 / 0.11\%} & \{3,4,5,6,8\},\{3,4,5,6,9\},\{3,4,5,6,10\},\{3,4,5,6,11\},\{3,4,14,15,16\}, \\ 
					~ & ~ & \{3,5,6,8,9\},\{3,5,6,8,10\},\{3,5,6,8,11\},\{3,5,6,9,10\} \\ 
					~ & ~ & \{3,5,6,9,11\},\{3,5,6,10,11\},\{8,9,14,15,16\},\{8,10,14,15,16\}, \\ 
					~ & ~ & \{8,11,14,15,16\},\{9,10,14,15,16\},\{9,11,14,15,16\}, \\ 
					~ & ~ & \{10,11,14,15,16\} \\ \hline 
					
				\end{tabular}
			} 
			
		\end{table}
		
		\subsection{Results on the RBTS system}
		
		RBTS system is a small-scale composite system with
		11 generations, 9 transmission lines and 6 buses. The total
		installed capacity of generations is 240 MW, with an annual peak load
		of 185MW.
		
		\subsubsection{Results of Critical States}
		For the entire system, the CSILP method partitioned out a total of 2,896 failure lattices from the state space and identified 62 critical states. These 62 critical states correspond to the minimal elements of 62 distinct failure lattices among the 2,896 lattices. The identified critical states are distributed across levels one to five, as shown in \autoref{FCS_RBTS}. The table also lists the number of critical states at each level and the proportion of these critical states relative to the total number of system states at each level. The proportion of critical states at level 2 is the highest, which demonstrates that identifying critical states in the 2-level state space of every lattice is crucial and effective.

		\begin{table*}[htbp]
			\centering
			\caption{Indices of Critical States(RBTS)}
			\label{RBTS_CS}
				\begin{tabular}{ccccccc}
					\toprule
					\multirow{2}*{Critical State} & \multirow{2}*{Level} & \multirow{2}*{Risk} & \multicolumn{2}{c}{$\Delta \text{LOLP}$} & \multirow{2}*{LOLP(\%)} & \multirow{2}*{OPF Number}  \\
					
					\cline{4-5}
					& & & Value (\%) & Proportion (\%) & & \\ 
					\midrule
					\(\{20\}\) & 1 & 1.81e-2 & 0.114155 & 12.047 & 0.1141553 & 20 \\
					\midrule
					\(\{1,2\}\) & 2 & 1.90e-2 & 0.089897 & 9.488 & 0.2040525 & 21 \\
					\(\{1,4\}\) & 2 & 3.14e-3 & 0.072667 & 7.669 & 0.2767195 & 23 \\
					\multicolumn{7}{c}{\(\vdots\)} 
					\\
					\(\{16,19\}\) & 2 & 4.14e-5 & 1.291e-4 & 0.013 & 0.9432331 & 188
					\\
					\midrule
					\(\{4,8,9\}\) & 3 & 2.36e-5 & 5.181e-4 & 0.055 & 0.9437511 & 559 \\
					\multicolumn{7}{c}{\(\vdots\)} 
					\\
					\(\{14,15,19\}\) & 3 & 8.91e-8 & 7.359e-7 & 7.8e-5 & 0.9475025 & 892
					\\
					\midrule
					\(\{1,14,15,16\}\) & 4 & 2.76e-9 & 1.942e-8 & 2.1e-6 & 0.9475025 & 930 \\
					\multicolumn{7}{c}{\(\vdots\)} 
					\\
					\(\{13,15,18,19\}\) & 4 & 1.36e-9 & 4.176e-9 & 4.41e-7 & 0.9475164 & 2901
					\\
					\midrule
					\(\{3,4,14,15,16\}\) & 5 & 1.55e-11 & 3.42e-10 & 3.61e-8 & 0.9475164 & 3677 \\
					\multicolumn{7}{c}{\(\vdots\)} \\
					\(\{3,5,6,10,11\}\) & 5 & 1.91e-9 & 3.921e-8 & 4.14e-6 & 0.9475169 & 4712 \\
					\bottomrule
				\end{tabular}
			
		\end{table*}

		As shown in \autoref{RBTS_CS}, some of the critical states and their associated data are listed in the order they were identified. The sixth and seventh columns represent the  LOLP and OPF number when the corresponding critical state in the first column was identified. Risk of a state was the product of probability and the loadshedding of the state according to \cite{ding2016bi_level}. \(\Delta \text{LOLP}\) of a state represents its total contribution to LOLP while \(\Delta \text{LOLP}\) Proportion represents the proportion of \(\Delta \text{LOLP}\) in the final LOLP(0.9475169\%). \(\Delta \text{LOLP}\) of these critical states generally show a downward trend in descending order. We can observe  that the order in which the critical states are identified by the CSILP method often reflects their risk levels and contributions to LOLP. Specifically, the critical states identified first tend to be those with higher risk and greater \(\Delta \text{LOLP}\). Therefore, this test system demonstrates that the proposed method can identify critical states in a reasonable order. However, there are some exceptions where the order deviates slightly from this trend, this is because the order of 1-level states in a 1-normal lattice to be partitioned is not optimal choice and there is still room for improvement.

		\subsubsection{Results of Reliability Assessment}
		For comparison, SE method traversed all possible system states of the entire system, and its assessment results serve as the baseline for calculating the accuracy of the other two methods; and the convergence criterion for the MCS method is set to a coefficient of variation of 0.01. 
		
\begin{figure}[htbp]  
			\centering
			\includegraphics[width=\columnwidth]{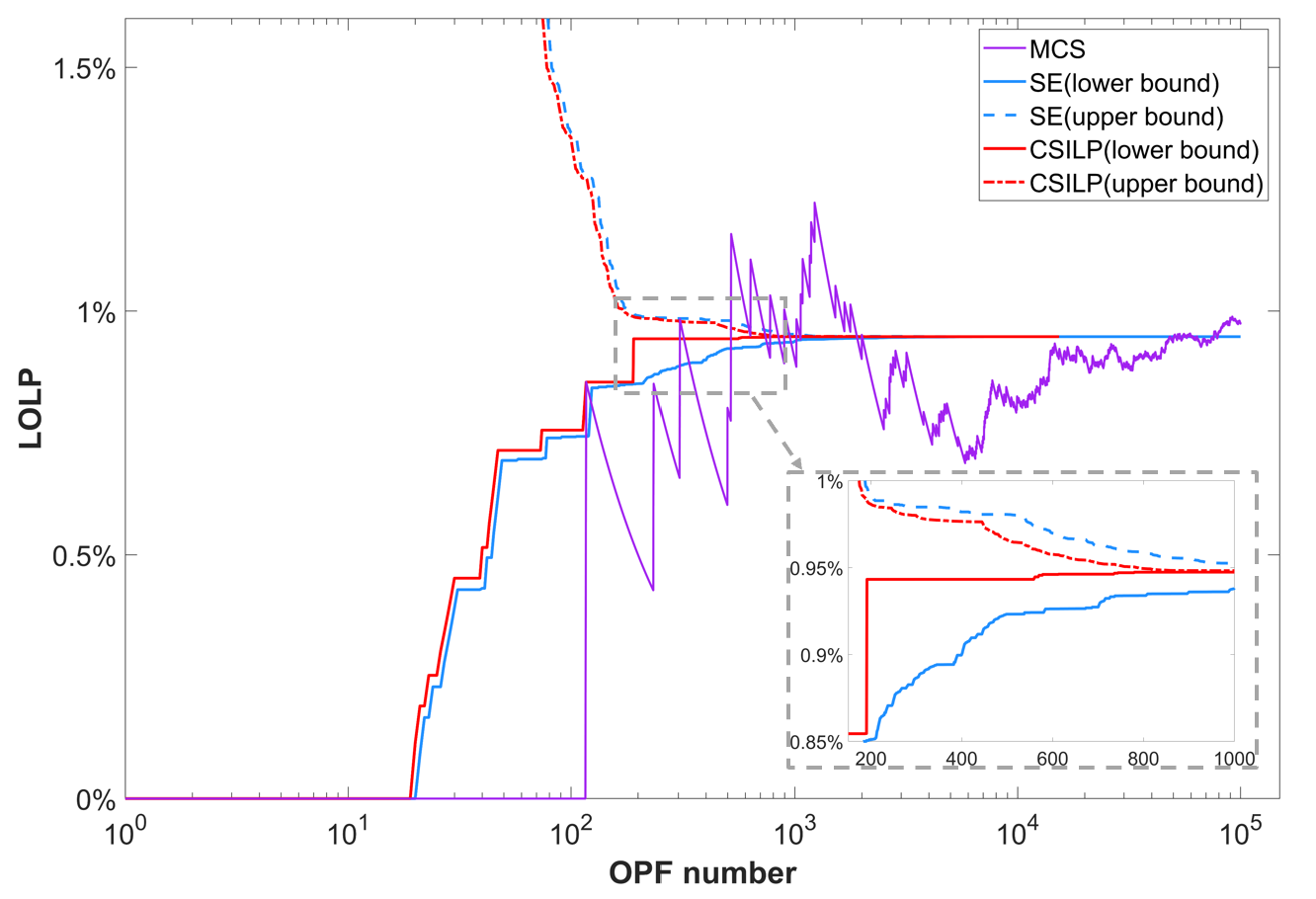}
			\caption{Results of three assessment methods (RBTS)}
			\label{experi:RBTS_LOLP&num2}
		\end{figure}

		As shown in \autoref{experi:RBTS_LOLP}, index yielded by SE and CSILP are exactly the same, so the error of CSILP is 0\%. However, error of MCS is 0.26\%. In addition, it cost SE method about 65,734 seconds to reach the analytical value. Such precision can be reached by CSILP within 887 seconds, which is almost 100 times faster than SE. MCS can not reach the precision even with more than 1000 seconds, which is slower and less efficiency. Convergence curves based on OPF number of MCS and results of CSILP and SE are also shown in Fig.\ref{experi:RBTS_LOLP&num2}, indicating that CSILP converges faster than SE and MCS, and the convergence of the MCS method is not only slow but also results in inaccurate values.

		\begin{table}[htbp]
			\centering
			\caption{Results of three assessment methods (RBTS)}
			\label{experi:RBTS_LOLP} 
				\begin{tabular}{ccccc} 
					\toprule[1pt] 
					\multirow{2}*{Method} & \multicolumn{2}{c}{LOLP} & \multirow{2}*{\makecell{OPF \\ Number}} & \multirow{2}*{\makecell{CPU \\ Time (s)}} \\ 
					\cline{2-3}
					& Value (\%) & Error (\%) & &\\ 
					\midrule 
					SE & 0.9475169361176 & —— & 1,048,576 & 65,734 \\ 
					MCS & 0.9053411493810 & 0.2648 & 19,882 & 1,200 \\ 
					CSILP & 0.9475169361176 & 0.0000 & 15,335 & 887 \\ 
					\bottomrule[1pt] 
				\end{tabular}
		\end{table}

		As shown in \autoref{expri:RBTS_OPF}, larger OPF number leads to a higher precision and a lower efficiency. However, when $N_*$ is identical, results yielded by CSILP are more accurate and time saving than SE. Specifically, the error of LOLP calculated by the CSILP method is nearly 0.45\% after 200 OPF calculations, which is a small and acceptable error range. This is because within 200 OPF calculations, the CSILP method has already evaluated all 1-level and 2-level states, which significantly contribute to the LOLP by covering the probabilities of many failure lattices. In contrast, the SE method requires more than 1000 OPF calculations to achieve a similar level of precision. And the index calculated by MCS are unstable, and even after a large number of OPF calculations, e.g. 15000 OPF calculations, the relative error remains still large.

		\begin{table}[htbp]
			\centering
			\caption{The impact of different convergence condition(OPF number) on accuracy and efficiency (RBTS)}
			\label{expri:RBTS_OPF}
				\begin{tabular}{cccccc}
					\hline
					\toprule[1pt]
					\multirow{2}*{$N_*$} & \multirow{2}*{Method} & \multicolumn{2}{c}{LOLP} & \multirow{2}*{$\overline{\text{LOLP}}$(\%)} & \multirow{2}*{\makecell{ CPU \\Time(s)}} \\
					\cline{3-4}
					& & Value (\%) & Error (\%) & & \\
					\hline
					\multirow{3}{*}{200} & MCS & 0.500000 & 47.2305 & —— & 10 \\ 
					~ & SE & 0.851036 & 10.1826 & 0.990620 & 10 \\ 
					~ & CSILP & 0.943224 & 0.4531 & 0.986330 & 17 \\ 
					\hline
					\multirow{3}{*}{1000} & MCS & 0.900000 & 5.0149 & —— & 80 \\ 
					~ & SE & 0.938204 & 0.9828 & 0.952506 & 66 \\ 
					~ & CSILP & 0.947390 & 0.0134 & 0.948216 & 75 \\ 
					\hline
					\multirow{3}{*}{5000} & MCS & 0.780000 & 17.6796 & —— & 301 \\ 
					~ & SE & 0.947189 & 0.0346 & 0.947570 & 273 \\ 
					~ & CSILP & 0.947517 & 2.65e-05 & 0.947520 & 307 \\ 
					\hline
					\multirow{3}{*}{15000} & MCS & 0.913333 & 3.6077 & —— & 899 \\ 
					~ & SE & 0.947497 & 0.0021 & 0.947520 & 787 \\ 
					~ & CSILP & 0.947517 & 2.38e-12 & 0.947517 & 868 \\ 
					\hline
				\end{tabular}
			
		\end{table}

		Furthermore, the preset gap between the upper and lower bounds of LOLP, $\delta_*$, also significantly impacts the performance of the CSILP method: a smaller $\delta_*$ requires more state evaluations, which we can observe from \autoref{expri:RBTS_theta}. For example, achieving \( \delta < 10^{-10} \) took 580 seconds and 9879 OPF calculations, and the relative error was reduced to \( 1.5 \times 10^{-7}\% \). In contrast, SE required nearly ten times more evaluations and time to reach the same $\delta_*$ and still resulted in a larger relative error.

		Besides, different values of level termination criteria $k_*$ were set as shown in \autoref{expri:RBTS_level}. $k_*$ can also affect the performance of CSILP. A larger $k_*$ 
		means a broader searching area and higher precision. We can see that the index calculated by CSILP are already very close to the true value, with a relative error of only \(6.45 \times 10^{-7}\%\) after evaluating all states up to the first five levels. This is because the critical states of the RBTS system are at most five levels high. Therefore, once all states up to the 5-level are evaluated, all critical states are identified, and the failure lattices generated by them cover the vast majority of the failure state space. However, SE needs to evaluate states up to the first seven levels to reach the same precision, and it took more than ten times the amount of time.

		\begin{table}[htbp]
			\centering
			\caption{The impact of different convergence condition (gap) on accuracy and efficiency (RBTS)}
			\label{expri:RBTS_theta}
			\resizebox{1\linewidth}{!}{
				\begin{tabular}{ccccccc}
					\hline
					\toprule[1pt]
					\multirow{2}*{$\delta_{*}$} & \multirow{2}*{Method} & \multicolumn{2}{c}{LOLP} & \multirow{2}*{$\overline{\text{LOLP}}$(\%)} & \multirow{2}*{\makecell{ OPF \\ Number} } & \multirow{2}*{\makecell{ CPU \\Time(s)}} \\ 
					\cline{3-4}
					& & Value (\%) & Error (\%) & & & \\ 
					\hline
					\multirow{2}{*}{$10^{-2}$} & SE & 0.696455 & 26.4968 & 1.690149 & 68 & 5 \\ 
					~ & CSILP & 0.714417 & 24.6011 & 0.017022 & 63 & 8 \\ 
					\hline
					\multirow{2}{*}{$10^{-6}$} & SE & 0.947425 & 0.00975 & 0.947525 & 8512 & 454 \\ 
					~ & CSILP & 0.947502 & 0.001559 & 0.947602 & 2382 & 163 \\ 
					\hline
					\multirow{2}{*}{$10^{-10}$} & SE & 0.947517 & 1.04e-06 & 0.947517 & 99395 & 5076 \\ 
					~ & CSILP & 0.947517 & 1.57e-07 & 0.947517 & 9879 & 580 \\ 
					\hline
				\end{tabular}
			}
			
		\end{table}

		\begin{table}[htbp]
			\centering
			\caption{The impact of different convergence condition (level) on accuracy and efficiency (RBTS)}
			\label{expri:RBTS_level} 
			\resizebox{1\linewidth}{!}{ 
				\begin{tabular}{ccccccc}
					\hline
					\multirow{2}*{$k_*$} & \multirow{2}*{Method} & \multicolumn{2}{c}{LOLP} & \multirow{2}*{$\overline{\text{LOLP}}$(\%)} & \multirow{2}*{\makecell{ OPF \\Number}} & \multirow{2}*{\makecell{ CPU \\Time(s)}} \\ 
					\cline{3-4}
					& & Value (\%) & Error (\%) & & &\\ 
					\hline
					\multirow{2}{*}{2} & SE & 0.852125 & 10.0676 & 0.988488 & 211 & 12 \\ 
					~ & CSILP & 0.943224 & 0.4531 & 0.988488 & 191 & 16 \\ \hline
					\multirow{2}{*}{3} & SE & 0.942294 & 0.5513 & 0.948266 & 1351 & 84 \\ 
					~ & CSILP & 0.947385 & 0.0139 & 0.948266 & 899 & 69 \\ \hline
					\multirow{2}{*}{4} & SE & 0.947336 & 0.0191 & 0.947525 & 6196 & 336 \\ 
					~ & CSILP & 0.947516 & 1.53e-04 & 0.947525 & 2905 & 191 \\ \hline
					\multirow{2}{*}{5} & SE & 0.947513 & 0.0005 & 0.947517 & 21700 & 1130 \\ 
					~ & CSILP & 0.947517 & 6.45e-07 & 0.947517 & 6658 & 399 \\ \hline
					\multirow{2}{*}{6} & SE & 0.947517 & 8.51e-06 & 0.947517 & 60460 & 3101 \\ 
					~ & CSILP & 0.947517 & 4.95e-09 & 0.947517 & 11123 & 649 \\ \hline
					\multicolumn{7}{c}{\vdots} \\
					\hline
					\multirow{2}{*}{9} & SE & 0.947517 & 1.87e-10 & 0.947517 & 431910 & 22081 \\ 
					~ & CSILP & 0.947517 & 0.0000 & 0.947517 & 15335 & 887 \\ \hline
				\end{tabular}
            }
		\end{table}


		\subsection{Results on RTS79 system}
		In this system, there are 24 buses, 32 generators and 38 transmission lines. The total installed capacity is 3405 MW, with an annual peak load of 2850 MW.

		\subsubsection{Results of Critical States}
		With $k_*$ in Algorithm2 set as 4, the proposed method partitioned out 4787 failure lattices from the the state space of RTS79 system and identified 2785 critical states of the first four levels. Some of the critical states of the first four levels are shown in \autoref{experi:RTS_Cdistribution} with the order they were identified, and there are no 1-level critical states in this state space since all the 1-level states are normal. The percentages indicate the proportion of these numbers relative to the total number of system states of that level. For example, the number of 2-level critical states is 15, accounting for 0.62\% of the total number of 2-level states. 
		Similar to \autoref{RBTS_CS}, \autoref{Indices of Critical States (RTS79)} lists some critical states and their related indices, the order in which these critical states were identified generally follows the order of their risk or \(\Delta \text{LOLP}\), from highest to lowest.
		

		\begin{table}[H]
			\centering
			\caption{Critical states (RTS79)} 
			\label{experi:RTS_Cdistribution} 
			\resizebox{1\linewidth}{!}{ 
				\begin{tabular}{ccc} 
					\toprule[1pt] 
					Level & \makecell{Number \\ / Proportion} & Critical State \\ \hline
					\multirow{2}{*}{2} & \multirow{2}{*}{15 / 0.62\%} & \{12,22\},\{12,23\},\{13,22\},\{13,23\},\{14,22\},\{14,23\},\{22,23\},\\
					~ & ~ & \{22,32\},\{22,43\},\{23,32\},\{23,43\},\{35,41\},\{36,40\},\{37,42\},\{51,55\} \\ \hline
					\multirow{4}{*}{3} & \multirow{4}{*}{383 / 0.7\%} & \{1,20,22\},\{1,21,22\},\{1,22,30\},\{1,22,31\},\{2,20,22\},\{2,21,22\},\\
					~ & ~ & \{2,22,30\},\{2,22,31\},\{3,9,22\},\{3,10,22\},\{3,11,22\},\{3,20,22\} \\ 
					~ & ~ & \(\vdots\) \\
					~ & ~ & \{55,59,61\},\{56,59,60\}\{57,58,60\},\{61,66,67\},\{61,68,69\} \\ \hline
					\multirow{4}{*}{4} & \multirow{4}{*}{2477 / 0.27\%} & \{20,22,44,45\},\{21,22,44,45\},\{22,30,44,45\},\{22,31,44,45\},\\
					~ & ~ & \{1,9,22,24\},\{1,9,22,25\},\{1,9,22,26\},\{1,9,22,27\},\{1,9,22,28\},\\
					~ & ~ & \(\vdots\) \\
					~ & ~ & \{56,57,58,59\},\{56,59,62,63\},\{57,58,62,63\},\{66,67,68,69\} \\ \hline
				\end{tabular}
			} 
			
		\end{table}

		\begin{table*}[htbp]
			\centering
			\caption{Indices of critical states(RTS79)}
			\label{Indices of Critical States (RTS79)}
				\begin{tabular}{ccccccc}
					\toprule
					\multirow{2}*{Critical State} & \multirow{2}*{Level} & \multirow{2}*{Risk} & \multicolumn{2}{c}{$\Delta \text{LOLP}$} & \multirow{2}*{LOLP(\%)} & \multirow{2}*{OPF Number}  \\ 
					\cline{4-5}
					& & & Value (\%) & proportion (\%) & &\\ 
					\midrule
					\(\{12,22\}\) & 2 & 6.95e-02 & 0.600000 & 7.130705 & 0.600000 & 784 \\
					\(\{12,23\}\) & 2 & 6.95e-02 & 0.528000 & 6.275020 & 1.128000 & 785 \\
					\(\{13,22\}\) & 2 & 6.95e-02 & 0.570000 & 6.774169 & 1.698000 & 841 \\
					\multicolumn{7}{c}{\vdots} \\
					\(\{51,55\}\) & 2 & 1.04e-05 & 2.157e-05 & 2.563e-04 & 5.906683 & 2299 \\
					\midrule
					\(\{1,20,22\}\) & 3 & 2.91e-03 & 0.033307 & 0.395836 & 5.939990 & 2501 \\
					\multicolumn{7}{c}{\vdots} \\
					\(\{61,68,69\}\) & 3 & 5.54e-09 & 4.531e-09 & 5.385e-08 & 8.282226 & 56149 \\
					\midrule
					\(\{20,22,44,45\}\) & 4 & 1.32e-09 & 3.379e-08 & 4.016e-07 & 8.282226 & 56548 \\
					\multicolumn{7}{c}{\(\vdots\)} \\
					\(\{66,67,68,69\}\) & 4 & 1.22e-12 & 9.57e-13 & 1.14e-11 & 8.414316 & 921223 \\
					\bottomrule
				\end{tabular}
		\end{table*}

		\subsubsection{Results of Reliability Assessment}
		
\begin{figure}[htbp]  
			\centering
			\includegraphics[width=\columnwidth]{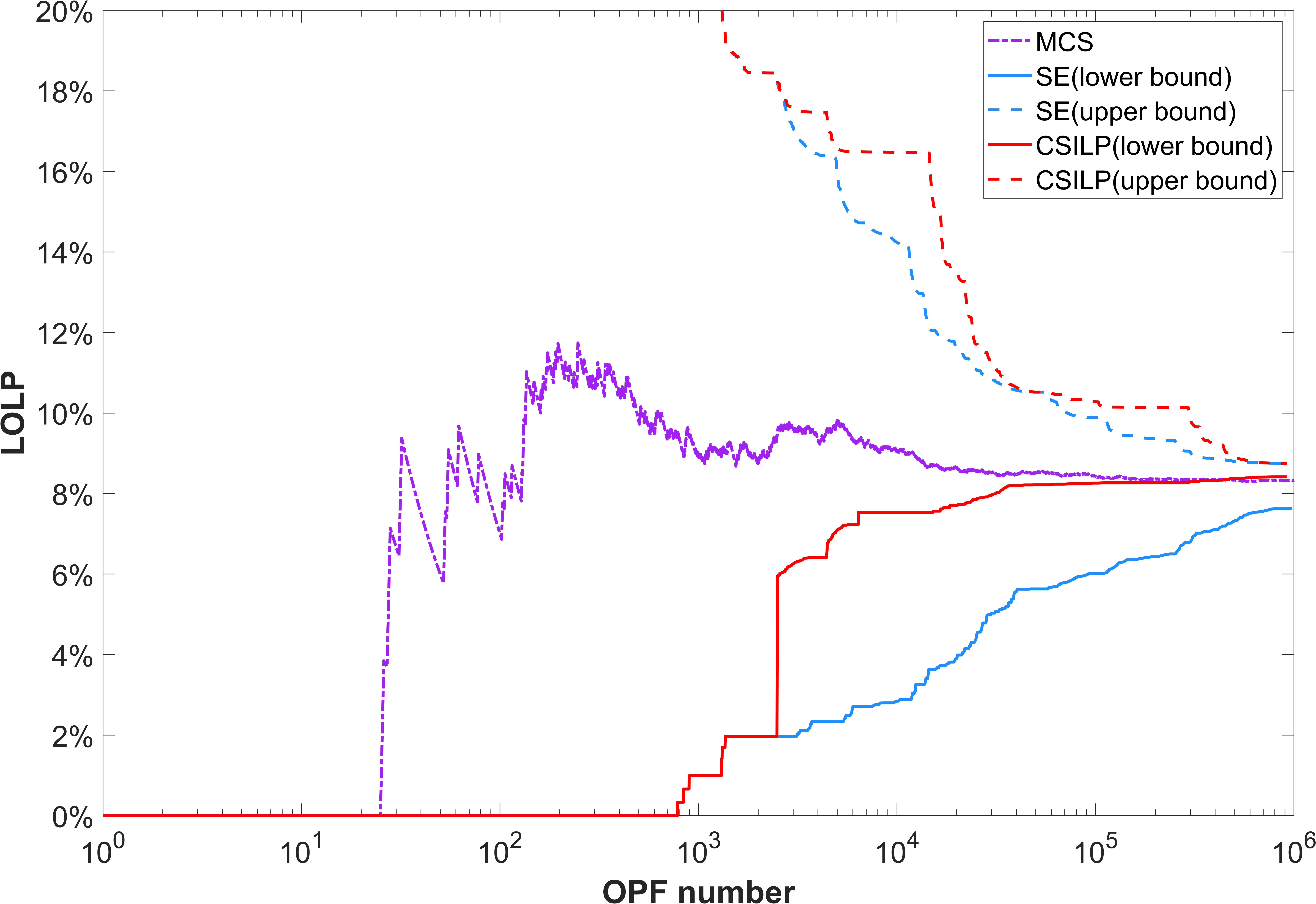}
			\caption{Results of three assessment methods(RTS79)}
			\label{expri:RTS_LOLP&OPF}
		\end{figure}

		CSILP method was applied to the system up to the forth level to test its accuracy and efficiency in reliability assessment. It was compared with SE and MCS, as shown in \autoref{expri:RTS_LOLP} and Fig.\ref{expri:RTS_LOLP&OPF}. SE enumerated all the states of the first four levels, and the convergence criterion for MCS was set as a preset total sampled state number of 1,000,000(finally reached variance of 0.004546). 
		
		The LOLP(\(\underline{\text{LOLP}}\)) and \(\overline{\text{LOLP}}\) of the RTS79 system generated by the CSILP method reached approximately 8.41\% and 8.75\% respectively, with only a 0.34\% gap between them. In contrast, the upper and lower bounds of LOLP generated by the SE exhibited a gap of approximately 1.13\%, which is relatively larger. Moreover, the CSILP method calculates LOLP values in a monotonically increasing manner, whereas the MCS method is inherently unstable. The LOLP value obtained by CSILP is higher than that calculated by the MCS method (8.33\%), which suggests that the CSILP method yields a more accurate result. Additionally, the CSILP method achieved these results with the fewest OPF calculations (921,157) among the three methods, which demonstrates that CSILP requires the least computational effort and provide the most accurate index. As shown in Fig.\ref{expri:RTS_LOLP&OPF}, the LOLP upper and lower bounds curves generated by the CSILP method converge more rapidly than those produced by the SE method and MCS method.

		\begin{table}[htbp]
			\centering
			\caption{Results of three assessment methods (RTS79)}
			\label{expri:RTS_LOLP} 
			\resizebox{1\linewidth}{!}{ 
				\begin{tabular}{ccccc} 
					\toprule[1pt] 
					Method & LOLP (\%) & $\overline{\text{LOLP}}$ (\%) &\makecell{OPF \\ Number} & \makecell{CPU \\ Time (s)} \\ 
					\midrule 
					SE & 7.620934 & 8.753119  & 974,120 & 60,214 \\ 
					MCS & 8.333100 & —— & 1,000,000 & 60,797 \\ 
					CSILP & 8.414316  & 8.753119 & 921,157 & 57,737 \\  
					\bottomrule[1pt] 
				\end{tabular}
			} 
		\end{table}
		
		LOLP generated by CSILP reaches a steady growth rate faster than SE as shown in \autoref{expri:RTS_OPFnum} and \autoref{expri:RTS_LOLP&OPF}. When $N_*$ is set 10,000, the LOLP calculated by CSILP reaches approximately 7.53\%. In contrast, it took SE more than 100 times the computational time and OPF number to achieve such precision. Meanwhile, the value calculated by MCS are unstable, sometimes even exceeding \(\overline{\text{LOLP}}\) calculated by CSILP. After 900,000 OPF calculations, the index calculated by MCS is 8.32\%, which is still less accurate than that calculated by CSILP. 
		
			\begin{table}[htbp]
			\centering
			\caption{The impact of different convergence condition(OPF number) on accuracy and efficiency (RTS79)}
			\label{expri:RTS_OPFnum}
			\resizebox{1\linewidth}{!}{
				\begin{tabular}{ccccc}
					\toprule
					$N_*$ & Method & LOLP(\%)  & \(\overline{\text{LOLP}}\)(\%) & CPU Time(s) \\ 
					\midrule
					\multirow{3}{*}{10000} & MCS & 9.070000 & —— & 753 \\ 
					& SE & 2.844094 & 14.233543 & 566 \\ 
					& CSILP & 7.528947 & 7.528947 & 882 \\ 
					\midrule
					\multirow{3}{*}{50000} & MCS & 8.520000 & —— & 4192 \\ 
					& SE & 5.629575 & 10.518847 & 2835 \\ 
					& CSILP & 8.216600 & 8.222141 & 3831 \\ 
					\midrule
					\multirow{3}{*}{200000} & MCS & 8.354000 & —— & 12987 \\ 
					& SE & 6.429055 & 9.370443 & 11444 \\ 
					& CSILP & 8.266879 & 8.292001 & 14602 \\ 
					\midrule
					\multirow{3}{*}{900000} & MCS & 8.323222 & —— & 54674 \\ 
					& SE & 7.620934 & 8.753119 & 54871 \\ 
					& CSILP & 8.414316 & 8.414316 & 56401 \\ 
					\bottomrule
				\end{tabular}
			}
		\end{table}

		As shown in \autoref{expri:RTS_gap},  under the same $\delta_*$, SE always requires a higher number of OPF calculations and more time compared to CSILP. Moreover, \(\overline{\text{LOLP}}\) of CSILP is always higher than that of SE. And SE can only achieve a maximum gap of 0.02 between the upper and lower bounds, whereas CSILP can achieve a gap that is an order of magnitude smaller(e.g. 0.004).

		The higher the $k_*$ is set, the more accurate the index become. As shown in \autoref{expri:RTS_level}, CSILP always achieves better index than SE at the same Level. And when \(k_*=1\) and 2, the OPF number required by SE and CSILP are almost the same. However, LOLP obtained by CSILP is more accurate than that of SE. This is because CSILP covers the probability of the failure lattices generated by 2-level failure states. The $\overline{\text{LOLP}}$ of the two methods in \autoref{expri:RTS_level} are almost the same, this is because the calculation of \(\overline{\text{LOLP}}\) depends on the sum of the probabilities of normal states, and both SE and CSILP methods handle the same set of normal states for each level evaluation.

		\begin{table}[H]
			\centering
			\caption{The impact of different convergence condition (gap) on accuracy and efficiency (RTS79)}
			\label{expri:RTS_gap}
			\resizebox{1\linewidth}{!}{
				\begin{tabular}{cccccc}
					\toprule
					$\delta_*$ & Method & LOLP(\%) & $\overline{\text{LOLP}}$(\%) & \makecell{OPF \\ Number} & \makecell{CPU \\ Time(s)} \\ 
					\midrule
					\multirow{2}{*}{0.1} & SE & 0.314348 & 13.099582 & 12370 & 701 \\ 
					& CSILP & 6.871609 & 16.870452 & 4680 & 428 \\ 
					\midrule
					\multirow{2}{*}{0.02} & SE & 6.938624 & 8.937154 & 309905 & 17876 \\ 
					& CSILP & 8.259270 & 10.258888 & 103519 & 8384 \\ 
					\midrule
					0.01 & CSILP & 8.324125 & 9.324124 & 352881 & 23340 \\ 
					\midrule
					0.004 & CSILP & 8.395841 & 8.795827 & 613184 & 38677 \\ 
					\bottomrule
				\end{tabular}
			}
		\end{table}

		\begin{table}[H]
			\caption{The impact of different convergence condition (level) on accuracy and efficiency (RTS79)}
			\label{expri:RTS_level}
			\centering
			\resizebox{1\linewidth}{!}{
				\begin{tabular}{cccccc}
					\toprule
					$k_*$ & Method & LOLP(\%) & $\overline{\text{LOLP}}$(\%) & \makecell{OPF \\ Number} & \makecell{CPU \\ Time(s)}  \\ 
					\midrule
					\multirow{2}{*}{1} & SE & 0.000000 & 41.845996 & 70 & 4 \\
					& CSILP & 0.000000 & 41.845996 & 70 & 9 \\
					\midrule
					\multirow{2}{*}{2} & SE & 1.969654 & 18.444269 & 2485 & 140 \\
					& CSILP & 5.906663 & 18.444269 & 2485 & 242 \\
					\midrule
					\multirow{2}{*}{3} & SE & 5.629576 & 10.518798 & 57225 & 3244 \\
					& CSILP & 8.216603 & 10.518798 & 56235 & 4364 \\
					\midrule
					\multirow{2}{*}{4} & SE & 7.620934 & 8.753119 & 974120 & 60214 \\
					& CSILP & 8.414316 & 8.753119 & 921227 & 57742 \\
					\bottomrule
				\end{tabular}
			}
			
		\end{table}

		
\section{Conclusion}			

To identify the critical states of a power system, this paper applied a lattice structure to represent and partition the state space of system, which enables the extension of failure states. In addition, based on lattice partition, a recursive method CSILP was proposed to identify all critical states up to a preset level and calculate the probability of all failure lattices during the process in an efficient way. 
Identifying critical states and calculating reliability index in real-time can be performed simultaneously. However, the growth rate of the index and the speed at which high-risk critical states are identified depend on the order of lattice partitioning and the criteria for selecting specific lattice to partition. The proposed method thus has significant room for improvement. How to choose a better order for partitioning will be a subject for future research.

 
\bibliographystyle{IEEEtran} 
\nocite{*}
\bibliography{ref}

\begin{thebibliography}{10}
\providecommand{\url}[1]{#1}
\csname url@samestyle\endcsname
\providecommand{\newblock}{\relax}
\providecommand{\bibinfo}[2]{#2}
\providecommand{\BIBentrySTDinterwordspacing}{\spaceskip=0pt\relax}
\providecommand{\BIBentryALTinterwordstretchfactor}{4}
\providecommand{\BIBentryALTinterwordspacing}{\spaceskip=\fontdimen2\font plus
\BIBentryALTinterwordstretchfactor\fontdimen3\font minus
  \fontdimen4\font\relax}
\providecommand{\BIBforeignlanguage}[2]{{%
\expandafter\ifx\csname l@#1\endcsname\relax
\typeout{** WARNING: IEEEtran.bst: No hyphenation pattern has been}%
\typeout{** loaded for the language `#1'. Using the pattern for}%
\typeout{** the default language instead.}%
\else
\language=\csname l@#1\endcsname
\fi
#2}}
\providecommand{\BIBdecl}{\relax}
\BIBdecl

\bibitem{Billinton1994}
R.~Billinton and W.~Li, \emph{Reliability Assessment of Electric Power Systems
  Using Monte Carlo Methods}.\hskip 1em plus 0.5em minus 0.4em\relax New York:
  Springer New York, NY, 1994.

\bibitem{lee2022reliability}
J.~W. Lee and S.~W. Kim, ``Reliability-centered maintenance strategy for
  redundant power networks using the cut set method,'' \emph{Journal of
  Electrical Engineering and Technology}, vol.~17, no.~3, pp. 1615--1621, 2022.

\bibitem{billinton1970power}
R.~Billinton, \emph{Power system reliability evaluation}.\hskip 1em plus 0.5em
  minus 0.4em\relax Taylor \& Francis, 1970.

\bibitem{Ford_Fulkerson_1956}
L.~R. Ford and D.~R. Fulkerson, ``Maximal flow through a network,''
  \emph{Canadian Journal of Mathematics}, vol.~8, p. 399–404, 1956.

\bibitem{10.5555/1942094}
D.~R. Ford and D.~R. Fulkerson, \emph{Flows in Networks}.\hskip 1em plus 0.5em
  minus 0.4em\relax USA: Princeton University Press, 2010.

\bibitem{billinton1992reliability}
R.~Billinton and R.~N. Allan, \emph{Reliability evaluation of engineering
  systems}.\hskip 1em plus 0.5em minus 0.4em\relax Springer, 1992, vol. 792.

\bibitem{5342441}
Y.~Liu and C.~Singh, ``Reliability evaluation of composite power systems using
  markov cut-set method,'' \emph{IEEE Transactions on Power Systems}, vol.~25,
  no.~2, pp. 777--785, 2010.

\bibitem{LIU20081019}
H.~Liu, Y.~Sun, P.~Wang, L.~Cheng, and L.~Goel, ``A novel state selection
  technique for power system reliability evaluation,'' \emph{Electric Power
  Systems Research}, vol.~78, no.~6, pp. 1019--1027, 2008.

\bibitem{6523187}
Y.~Jia, P.~Wang, X.~Han, J.~Tian, and C.~Singh, ``A fast contingency screening
  technique for generation system reliability evaluation,'' \emph{IEEE
  Transactions on Power Systems}, vol.~28, no.~4, pp. 4127--4133, 2013.

\bibitem{ding2016bi_level}
T.~Ding, C.~Li, C.~Yan, and F.~F. Li, ``A bi-level optimization model for risk
  assessment and contingency ranking in transmission system reliability
  evaluation,'' \emph{IEEE Transactions on Power Systems}, vol.~31, no.~6, pp.
  1--1, December 2016.

\bibitem{CLANCY1983101}
D.~Clancy, G.~Gross, and F.~Wu, ``Probabilitic flows for reliability evaluation
  of multiarea power system interconnections,'' \emph{International Journal of
  Electrical Power and Energy Systems}, vol.~5, no.~2, pp. 101--114, 1983.

\bibitem{BILLINTON1998189}
R.~Billinton and W.~Zhang, ``State extension in adequacy evaluation of
  composite power systems—concept and algorithm,'' \emph{Electric Power
  Systems Research}, vol.~47, no.~3, pp. 189--195, 1998.

\bibitem{852155}
------, ``State extension for adequacy evaluation of composite power
  systems-applications,'' \emph{IEEE Transactions on Power Systems}, vol.~15,
  no.~1, pp. 427--432, 2000.

\bibitem{7741639}
K.~Hou, H.~Jia, X.~Yu, L.~Zhu, X.~Xu, and X.~Li, ``An impact increments-based
  state enumeration reliability assessment approach and its application in
  transmission systems,'' in \emph{2016 IEEE Power and Energy Society General
  Meeting (PESGM)}, 2016, pp. 1--5.

\bibitem{sagan2020combinatorics}
B.~E. Sagan, \emph{Combinatorics: The art of counting}.\hskip 1em plus 0.5em
  minus 0.4em\relax American Mathematical Soc., 2020, vol. 210.

\bibitem{doi:10.1049/iet-gtd.2009.0281}
J.~He, Y.~Sun, D.~Kirschen, C.~Singh, and L.~Cheng, ``State-space partitioning
  method for composite power system reliability assessment,'' \emph{IET
  Generation, Transmission and Distribution}, vol.~4, pp. 780--792, 2010.

\bibitem{159800}
J.~Dugan, S.~Bavuso, and M.~Boyd, ``Dynamic fault-tree models for
  fault-tolerant computer systems,'' \emph{IEEE Transactions on Reliability},
  vol.~41, no.~3, pp. 363--377, 1992.

\bibitem{KUMAR20182521}
T.~B. Kumar, O.~C. Sekhar, and M.~Ramamoorty, ``Composite power system
  reliability evaluation using modified minimal cut set approach,''
  \emph{Alexandria Engineering Journal}, vol.~57, no.~4, pp. 2521--2528, 2018.

\bibitem{6246198}
Z.~Zhou, Z.~Gong, B.~Zeng, L.~He, and D.~Ling, ``Reliability analysis of
  distribution system based on the minimum cut-set method,'' in \emph{2012
  International Conference on Quality, Reliability, Risk, Maintenance, and
  Safety Engineering}, 2012, pp. 112--116.

\bibitem{7131786}
B.~Lami and K.~Bhattacharya, ``Identification of critical components of
  composite power systems using minimal cut sets,'' in \emph{2015 IEEE Power
  and Energy Society Innovative Smart Grid Technologies Conference (ISGT)},
  2015, pp. 1--5.

\bibitem{6039271}
Y.~Jia, Z.~Yan, and P.~Wang, ``An improved state selection technique for power
  system reliability evaluation,'' in \emph{2011 IEEE Power and Energy Society
  General Meeting}, 2011, pp. 1--6.

\end{thebibliography}

\end{document}